\documentclass[11pt]{article}
\usepackage{epsfig} 
\newcommand{\sect}[1]{ \section{#1} \setcounter{equation}{0} }

\newcommand{\pslash}{p \! \! \! /}

\newcommand{\half}{\mbox{\small{$\frac{1}{2}$}}} 
 
\newcommand{\Nf}{N_{\!f}} 
\newcommand{\MSbar}{\overline{\mbox{MS}}}

\begin{document}
\title{Three loop anomalous dimension of the second moment of the transversity
operator in the $\MSbar$ and RI${}^\prime$ schemes} 
\author{J.A. Gracey, \\ Theoretical Physics Division, \\ 
Department of Mathematical Sciences, \\ University of Liverpool, \\ P.O. Box 
147, \\ Liverpool, \\ L69 3BX, \\ United Kingdom.} 
\date{} 
\maketitle 
\vspace{5cm} 
\noindent 
{\bf Abstract.} We compute the anomalous dimension of the second moment of the
transversity operator, $\bar{\psi} \sigma^{\mu\nu} D^\rho \psi$, at three loops
in both the $\MSbar$ and RI$^\prime$ schemes. As a check on the result we also 
determine the $O(1/\Nf)$ critical exponent of the $n$-th moment of the 
transversity operator in $d$-dimensions in the large $\Nf$ expansion and 
determine leading order information on the $n$ dependence of the anomalous 
dimension at three and four loops in $\MSbar$. In addition the RI$^\prime$ 
anomalous dimension of the non-singlet twist-$2$ operator, $\bar{\psi} 
\gamma^\mu D^\nu \psi$, is also determined.

\vspace{-17cm}
\hspace{13.5cm}
{\bf LTH 580} 

\newpage

\sect{Introduction.} 

Recently there has been renewed interest in the transversity distribution of 
the partons in the nucleons. Introduced originally in \cite{1,2,3} the 
transversity measures the difference in probabilities of finding a quark in a 
transversely polarized nucleon which is polarized parallel to that of the 
nucleon with that in the antiparallel polarization. Although there has not been
as large an experimental activity in this area compared with the conventional 
distribution functions of deep inelastic scattering, it is possible RHIC may
take transversity data in the near future, \cite{4}. (For recent reviews see, 
for instance, \cite{5,6}.) Therefore, to improve our interpretation of the 
results it is necessary to extend our theoretical understanding to a similar 
level as that achieved in deep inelastic scattering. For instance, there has 
been a large activity in determining the three loop anomalous dimensions of the
twist-$2$ unpolarized and polarized flavour non-singlet and singlet operators 
as a function of the operator moment, $n$, \cite{7,8,9,10,11}. These three loop 
results are necessary for the full two loop renormalization group evolution of 
the structure functions. Exact analytic results as a function of $n$ are 
required since the Mellin transform with respect to $n$ determine the parton 
splitting functions as a function of the momentum fraction, $x$. Such three 
loop calculations, however, require a huge increase in resources compared with 
the earlier one and two loop results of \cite{12,13,14,15,16,17,18,19}. For 
instance, due to the presence of a large number of Feynman diagrams with and 
without subgraphs, one encounters nested or harmonic sums whose mathematics has
had to be studied and developed in, for example, \cite{20}. Moreover, this has 
then to be encoded in the symbolic manipulation language {\sc Form}, \cite{21},
in order to handle the incredibly large amounts of algebra. Whilst results for 
non-singlet operators are expected soon it is clear that the same machinery can
be applied to determining the anomalous dimensions of the moments of the 
transversity operator as they differ only in their $\gamma$-algebra structure. 
For instance, the transversity operator involves the antisymmetric tensor 
$\sigma^{\mu\nu}$~$=$~$\half [ \gamma^\mu, \gamma^\nu ]$ instead of one 
$\gamma$-matrix. Whilst an analytic result for transversity at three loops is 
some way off, one can turn to the original three loop approach of \cite{7} and 
determine information on the low moments of the anomalous dimension of the 
operator. Indeed in \cite{22} the anomalous dimension of the first moment is 
available at three loops in QCD, building on the two loop result for this 
tensor current, \cite{23}. Moreover, the anomalous dimension of $\bar{\psi} 
\sigma^{\mu\nu} \psi$ is also available in QED at four loops in $\MSbar$ in the
quenched approximation, \cite{24}. In this article we will extend \cite{22} by 
computing the anomalous dimension of $\bar{\psi} \sigma^{\mu\nu} D^\rho \psi$ 
at three loops where $D^\mu$ is the usual covariant derivative. Although one 
motivation for the result lies in the provision of an independent calculation 
which will be useful to check against the result as a function of $n$ when it 
is determined, there are several other reasons for concentrating on this one 
operator. These lie in other theoretical techniques to determine information on
the associated matrix element. Using a lattice regularization one can deduce 
numerical estimates for matrix elements. However, such results are necessarily 
in a lattice regularization scheme and need to be matched to the standard 
$\MSbar$ renormalization scheme. Therefore, we will not only determine the 
anomalous dimension at three loops in $\MSbar$ but also in the RI$^\prime$ 
scheme which denotes the modified regularization invariant scheme, \cite{25}. 
This scheme was introduced in \cite{25} and developed for the problem of 
improving estimates for quark masses in \cite{26,27}. For applications to 
similar problems in applying the lattice to construct matrix elements relevant 
in deep inelastic scattering see, for instance, \cite{28,29}. It is also worth 
drawing attention to an alternative lattice approach to constructing moments 
of the parton distribution functions, \cite{30}. This uses the Schr\"{o}dinger 
functional technique of \cite{31} and is non-perturbative. More recently the 
RI$^\prime$ scheme has been examined in detail in \cite{32} where QCD has first
been renormalized at three loops before reproducing the results of \cite{27} at
three loops for the scalar current. Hence the RI$^\prime$ expression for the 
anomalous dimension of the tensor current was determined to the same order, 
\cite{32}. By contrast to $\MSbar$ the anomalous dimensions of these currents 
cease to be independent of the covariant gauge parameter. However, only results
are required in the Landau gauge for the lattice matching. Nevertheless from 
the point of view of internal consistency in multiloop computations we will 
perform our calculations in an arbitrary covariant gauge before deducing the 
requisite conversion functions in the Landau gauge. The disadvantage of an 
arbitrary gauge calculation is that the computation is algebraically more 
intense which is further complicated by the fact that the presence of a 
covariant derivative, itself, introduces another level of integration by parts,
also slowing the procedure. Therefore, we have chosen to only focus on the 
second moment. Choosing the Feynman gauge would speed the computer algebra 
manipulations but would not be useful for the necessary matching which is gauge
dependent. Further, matrix elements for higher moment operators from the 
lattice are currently much harder to extract. 

To be more specific about the results we will report here, we will renormalize
both the (flavour non-singlet) operators $\bar{\psi} \gamma^\mu D^\nu \psi$ and
$\bar{\psi} \sigma^{\mu\nu} D^\rho \psi$ at three loops in $\MSbar$ and 
RI$^\prime$. Although the anomalous dimension of the former operator is known 
in $\MSbar$, \cite{12,13,14,15,16,17,18,19,7}, we will use it as a check on the 
computer algebra programmes we have written in the symbolic manipulation 
language {\sc Form}, \cite{21}, before extracting the RI$^\prime$ anomalous
dimension. Then the Feynman rule for the operator insertion in a quark 
two-point function will be replaced with that for the transversity operator. 
One check on both the $\MSbar$ results will be that the anomalous dimensions of
both operators will be independent of the covariant gauge parameter, $\alpha$. 
The method of calculation is to apply the {\sc Mincer} algorithm, \cite{33}, 
written in {\sc Form}, \cite{34}, which determines the poles in $\epsilon$ in 
dimensional regularization, with $d$~$=$~$4$~$-$~$2\epsilon$, as well as the 
finite part of massless three loop two-point functions. Each operator will be 
inserted at zero external momentum to form a two-point function. Moreover, 
lattice calculations are for the operator inserted at zero momentum so the 
renormalization scheme conversion functions are required for this 
configuration. We have used {\sc Qgraf}, \cite{35}, to generate the Feynman 
diagrams to three loops and the output has been converted to the format used by
the {\sc Form} version of {\sc Mincer}.  

The paper is organised as follows. In section $2$ we introduce the method we
will use by considering the twist-$2$ flavour non-singlet operator and 
reproduce the three loop anomalous dimension for the second moment in $\MSbar$ 
before deriving the RI$^\prime$ scheme result. This approach is extended in 
section 3 to the case of the second moment of the transversity operator. The
conversion functions which allow one to translate from one renormalization 
scheme to another for each of the operators we consider are discussed in 
section $4$ and provide a partial check on our results. As an independent check
on our three loop result for the transversity we compute the large $\Nf$ 
leading order critical exponent for the $n$th moment of the operator using the
large $\Nf$ critical point technique in section $5$. Finally, we conclude with
discussion in section $6$. 

\sect{Second moment of non-singlet twist-$2$ operator.}

As we are considering operator insertions similar to the tensor current which
was studied in \cite{22,32}, we briefly recall the RI$^\prime$ renormalization
scheme definitions. To determine the correct finite part of the Green's
function with the operator inserted we must project out all possible components
consistent with Lorentz symmetry. For instance, for the flavour non-singlet
operator $\bar{\psi} \gamma^\mu D^\nu \psi$ we have in momentum space, 
\begin{eqnarray} 
G^{\mu\nu}_{\bar{\psi} \gamma^\mu D^\nu \psi} (p) &=&  
\langle \psi(p) ~ [ {\cal S} \bar{\psi} \gamma^\mu D^\nu \psi](0) ~ \bar{\psi}
(-p) \rangle \nonumber \\
&=& \Sigma^{(1)}_{\bar{\psi} \gamma^\mu D^\nu \psi}(p) \left( \gamma^\mu p^\nu
+ \gamma^\nu p^\mu - \frac{2}{d} \pslash \eta^{\mu\nu} \right) \nonumber \\
&& +~ \Sigma^{(2)}_{\bar{\psi} \gamma^\mu D^\nu \psi}(p) \left( p^\mu p^\nu 
\pslash - \frac{p^2}{d} \pslash \eta^{\mu\nu} \right) 
\end{eqnarray} 
where $p$ is the momentum flowing through the external quark legs and 
${\cal S}$ denotes both the symmetrization with respect to $\mu$ and $\nu$ and 
that the operator is traceless\footnote{The symmetrization and traceless 
symbol, ${\cal S}$, is understood in all our subscripts.}. We could have 
introduced the null vector, $\Delta_\mu$, to achieve this but it is not 
necessary here as we will in fact reproduce the known anomalous dimension for 
$\bar{\psi} \gamma^\mu D^\nu \psi$. To be more specifc  
\begin{equation} 
{\cal S} \bar{\psi} \gamma^\mu D^\nu \psi ~=~ \bar{\psi} \gamma^\mu D^\nu 
\psi ~+~ \bar{\psi} \gamma^\nu D^\mu \psi ~-~ \frac{2}{d} \eta^{\mu\nu} 
\bar{\psi} \gamma^\sigma D_\sigma \psi ~.  
\end{equation} 
The various components are determined through 
\begin{eqnarray} 
\Sigma^{(1)}_{\bar{\psi} \gamma^\mu D^\nu \psi}(p) &=& \frac{1}{8(d-1)} \left[ 
\mbox{tr} \left( \left( \gamma_\mu p_\nu + \gamma_\nu p_\mu - \frac{2}{d} 
\pslash \eta_{\mu\nu} \right) G^{\mu\nu}_{\bar{\psi} \gamma^\mu D^\nu \psi}(p) 
\right) \right. \nonumber \\
&& \left. ~~~~~~~~~~~~~-~ 2 \, \mbox{tr} \left( \left( p_\mu p_\nu \pslash 
- \frac{p^2}{d} \pslash \eta_{\mu\nu} \right) G^{\mu\nu}_{\bar{\psi} 
\gamma^\mu D^\nu \psi}(p) \right) \right] \nonumber \\  
\Sigma^{(2)}_{\bar{\psi} \gamma^\mu D^\nu \psi}(p) &=& -~ \frac{1}{4(d-1)} 
\left[ \mbox{tr} \left( \left( \gamma_\mu p_\nu + \gamma_\nu p_\mu 
- \frac{2}{d} \pslash \eta_{\mu\nu} \right) G^{\mu\nu}_{\bar{\psi} \gamma^\mu 
D^\nu \psi}(p) \right) \right. \nonumber \\
&& \left. ~~~~~~~~~~~~~~~~-~ (d+2) \, \mbox{tr} \left( \left( p_\mu p_\nu 
\pslash - \frac{p^2}{d} \pslash \eta_{\mu\nu} \right) G^{\mu\nu}_{\bar{\psi} 
\gamma^\mu D^\nu \psi}(p) \right) \right] ~.  
\end{eqnarray} 
To renormalize $\langle \psi(p) ~ [ {\cal S} \bar{\psi} \gamma^\mu D^\nu \psi]
(0) ~ \bar{\psi}(-p) \rangle$ in our symbolic manipulation approach we follow 
the method of \cite{36}. There the bare Green's function is computed in terms 
of the bare coupling constant and covariant gauge parameter. Renormalized 
variables are introduced through the usual renormalization constant definitions
such as $g_{\mbox{\footnotesize{o}}}$ $=$ $\mu^\epsilon Z_g g$ where $g$ is the 
coupling constant appearing in the covariant derivative $D_\mu$ $=$ 
$\partial_\mu$ $+$ $ig T^a A^a_\mu$, $\mu$ is the renormalization scale 
introduced to ensure the coupling constant remains dimensionless in 
$d$-dimensions and $T^a$ are the generators of the colour group whose structure
functions are $f^{abc}$. The renormalization constant associated with the 
operator, $Z_{\cal O}$, is deduced by ensuring that the Green's function is 
finite after multiplying by $Z_{\cal O} Z_\psi$ where the quark wave function 
renormalization constant, $Z_\psi$, arises from the external quark legs. For 
the RI$^\prime$ scheme it is the first component which determines 
$Z_{\cal O}^{\mbox{\footnotesize{RI$^\prime$}}}$ since this renormalization 
constant is defined by, \cite{25,32}, 
\begin{equation}  
\left. \lim_{\epsilon \, \rightarrow \, 0} \left[ 
Z^{\mbox{\footnotesize{RI$^\prime$}}}_\psi  
Z^{\mbox{\footnotesize{RI$^\prime$}}}_{\bar{\psi} \gamma^\mu D^\nu \psi}  
\Sigma^{(1)}_{\bar{\psi} \gamma^\mu D^\nu \psi}(p) \right] \right|_{p^2 \, = \, 
\mu^2} ~=~ 1 ~. 
\end{equation}  
In other words, unlike the $\MSbar$ scheme in addition to the poles in 
$\epsilon$ the $O(1)$ piece is absorbed into the renormalization constant. 
Though it is important to note that this renormalization constant will also
render the other components finite but not necessarily zero or unity.

Having summarized our method of calculation, we now record our results 
for the renormalization of $\bar{\psi} \gamma^\mu D^\nu \psi$. We find 
\begin{eqnarray} 
Z^{\mbox{\footnotesize{$\MSbar$}}}_{\bar{\psi} \gamma^\mu D^\nu \psi} &=& 1 ~+~
\frac{8}{3\epsilon} C_F a \nonumber \\
&& +~ \left[ \frac{4}{9} \left( 4 T_F \Nf + 8 C_F - 11 C_A \right) 
\frac{1}{\epsilon^2} ~+~ \frac{4}{27} \left( 47 C_A - 14 C_F - 16 T_F
\Nf \right) \frac{1}{\epsilon} \right] C_F a^2 \nonumber \\
&& +~ \left[ \frac{8}{81} \left( 121 C_A^2 - 132 C_A C_F - 88 C_A T_F \Nf 
+ 32 C_F^2 + 48 C_F T_F \Nf + 16 T_F^2 \Nf^2 \right) \frac{1}{\epsilon^3}
\right. \nonumber \\
&& \left. ~~~~+~ \frac{8}{243} \left( 718 C_A C_F - 823 C_A^2 + 544 C_A T_F \Nf
- 168 C_F^2 \right. \right. \nonumber \\
&& \left. \left. ~~~~~~~~~~~~~~~-~ 140 C_F T_F \Nf - 64 T_F^2 \Nf^2 \right) 
\frac{1}{\epsilon^2} \right. \nonumber \\  
&& \left. ~~~~+~ \frac{8}{729} \left( \left( 648 \zeta(3) + 2615 \right) C_A^2
- \left( 1944 \zeta(3) + 1066 \right) C_A C_F \right. \right. \nonumber \\
&& \left. \left. ~~~~~~~~~~~~~~~-~ \left( 1296 \zeta(3) + 782 \right) C_A T_F
\Nf + \left( 1296 \zeta(3) - 70 \right) C_F^2 \right. \right. \nonumber \\
&& \left. \left. ~~~~~~~~~~~~~~~+~ \left( 1296 \zeta(3) - 853 \right) C_F T_F 
\Nf - 112 T_F^2 \Nf^2 \right) \frac{1}{\epsilon} \right] C_F a^3 ~+~ O(a^4) 
\label{Zdis2msb} 
\end{eqnarray} 
which is clearly gauge independent and agrees exactly with the result of 
\cite{7}. Throughout we will present each renormalization constant as an aid to 
interested readers who wish to perform the check that the double and triple 
poles in $\epsilon$ can be derived from the simple poles of the previous loop 
order which follows from the fact that $\MSbar$ is a mass independent 
renormalization scheme. Further, in 
(\ref{Zdis2msb}) $\zeta(n)$ is the Riemann zeta function and we have set 
$a$~$=$~$g^2/(16\pi^2)$, $T^a T^a$~$=$~$C_F I$, 
$f^{acd} f^{bcd}$~$=$~$C_A \delta^{ab}$ and $\mbox{Tr} \left( T^a T^b 
\right)$~$=$~$T_F \delta^{ab}$. Consequently, having checked that our
programming reproduces known results it is elementary to change the 
renormalization scheme and deduce 
\begin{eqnarray} 
Z^{\mbox{\footnotesize{RI$^\prime$}}}_{\bar{\psi} \gamma^\mu D^\nu \psi} &=& 
1 ~+~ \left[ \frac{8}{3\epsilon} + \frac{1}{9} \left( 9 \alpha + 31 \right)
\right] C_F a \nonumber \\
&& +~ \left[ \frac{4}{9} \left( 4 T_F \Nf + 8 C_F - 11 C_A \right) 
\frac{1}{\epsilon^2} ~+~ \frac{4}{27} \left( 47 C_A + ( 48 + 18 \alpha ) C_F 
- 16 T_F \Nf \right) \frac{1}{\epsilon} \right. \nonumber \\
&& \left. ~~~~~+~ \frac{1}{324} \left( \left( 81 \alpha^3 + 486 \alpha^2 
- 324 \zeta(3) \alpha + 2439 \alpha - 3564 \zeta(3) + 12808 \right) C_A 
\right. \right. \nonumber \\
&& \left. \left. ~~~~~~~~~~~~~~~~~+ \left( 162 \alpha^2 + 630 \alpha 
+ 2592 \zeta(3) - 456 \right) \! C_F - \left( 720 \alpha + 5336 \right) T_F \Nf 
\right) \! \right] a^2 \nonumber \\ 
&& +~ \left[ \frac{8}{81} \left( 121 C_A^2 - 132 C_A C_F - 88 C_A T_F \Nf 
+ 32 C_F^2 + 48 C_F T_F \Nf + 16 T_F^2 \Nf^2 \right) \frac{1}{\epsilon^3}
\right. \nonumber \\
&& \left. ~~~~+~ \frac{4}{243} \left( ( 413 - 297 \alpha ) C_A C_F 
- 1646 C_A^2 + 1088 C_A T_F \Nf + ( 216 \alpha + 408 ) C_F^2 \right. \right. 
\nonumber \\
&& \left. \left. ~~~~~~~~~~~~~~~+~ ( 108 \alpha + 92 ) C_F T_F \Nf - 128 T_F^2 
\Nf^2 \right) \frac{1}{\epsilon^2} \right. \nonumber \\  
&& \left. ~~~~+~ \frac{2}{729} \left( \left( 243 \alpha^3 + 1458 \alpha^2 
- 972 \zeta(3) \alpha + 9855 \alpha - 18468 \zeta(3) + 42902 \right) C_A C_F 
\right. \right. \nonumber \\
&& \left. \left. ~~~~~~~~~~~~~~~+~ \left( 2592 \zeta(3) + 10460 \right) C_A^2 
- \left( 5184 \zeta(3) + 3182 \right) C_A T_F \Nf \right. \right. \nonumber \\
&& \left. \left. ~~~~~~~~~~~~~~~+~ \left( 486 \alpha^2 + 1134 \alpha - 4252 
+ 12960 \zeta(3) \right) C_F^2 - 448 T_F^2 \Nf^2 \right. \right. \nonumber \\
&& \left. \left. ~~~~~~~~~~~~~~~+~ \left( 5184 \zeta(3) - 22396 - 3024 \alpha 
\right) C_F T_F \Nf \right) \frac{1}{\epsilon} \right. \nonumber \\
&& \left. ~~~~+~ \frac{1}{69984} \left( \left( 8748 \alpha^5 + 83106 \alpha^4 
- 17496 \zeta(3) \alpha^3 + 493776 \alpha^3 + 1659204 \alpha^2 \right. \right.
\right. \nonumber \\
&& \left. \left. \left. ~~~~~~~~~~~~~~~~~~~-~ 96228 \zeta(3) \alpha^2
- 2523312 \zeta(3) \alpha + 174960 \zeta(5) \alpha + 7952229 \alpha \right. 
\right. \right. \nonumber \\
&& \left. \left. \left. ~~~~~~~~~~~~~~~~~~~-~ 11944044 \zeta(3) 
+ 746496 \zeta(4) + 524880 \zeta(5) + 38226589 \right) C_A^2 \right. \right.
\nonumber \\
&& \left. \left. ~~~~~~~~~~~~~~~~~~~+~ \left( 17496 \alpha^4 + 95256 \alpha^3 
+ 151632 \zeta(3) \alpha^2 + 416016 \alpha^2 \right. \right. \right. 
\nonumber \\
&& \left. \left. \left. ~~~~~~~~~~~~~~~~~~~~~~~~+~ 346032 \zeta(3) \alpha 
+ 466560 \zeta(5) \alpha + 921564 \alpha - 4914432 \zeta(3) \right. \right. 
\right. \nonumber \\
&& \left. \left. \left. ~~~~~~~~~~~~~~~~~~~~~~~~-~ 2239488 \zeta(4) 
+ 8864640 \zeta(5) + 3993332 \right) C_A C_F \right. \right. \nonumber \\
&& \left. \left. ~~~~~~~~~~~~~~~~~~~-~ \left( 77760 \alpha^3 + 466560 \alpha^2 
- 124416 \zeta(3) \alpha + 4091040 \alpha \right. \right. \right. \nonumber \\
&& \left. \left. \left. ~~~~~~~~~~~~~~~~~~~~~~~~-~ 369792 \zeta(3) 
+ 1492992 \zeta(4) + 24752896 \right) C_A T_F \Nf \right. \right. \nonumber \\
&& \left. \left. ~~~~~~~~~~~~~~~~~~~-~ \left( 155520 \alpha^2 - 622080 \zeta(3)
\alpha + 1937088 \alpha - 3234816 \zeta(3) \right. \right. \right. \nonumber \\
&& \left. \left. \left. ~~~~~~~~~~~~~~~~~~~~~~~~-~ 1492992 \zeta(4) 
+ 9980032 \right) C_F T_F \Nf \right. \right. \nonumber \\
&& \left. \left. ~~~~~~~~~~~~~~~~~~~+~ \left( 345600 \alpha + 221184 \zeta(3) 
+ 3391744 \right) T_F^2 \Nf^2 \right. \right. \nonumber \\ 
&& \left. \left. ~~~~~~~~~~~~~~~~~~~+~ \left( 17496 \alpha^3 + 289656 \alpha^2 
- 373248 \zeta(3) \alpha^2 - 715392 \zeta(3) \alpha \right. \right. \right.
\nonumber \\
&& \left. \left. \left. ~~~~~~~~~~~~~~~~~~~~~~~~+~ 879336 \alpha
+ 10737792 \zeta(3) + 1492992 \zeta(4) \right. \right. \right. \nonumber \\
&& \left. \left. \left. ~~~~~~~~~~~~~~~~~~~~~~~~-~ 9331200 \zeta(5) 
- 3848760 \right) C_F^2 \right) \right] C_F a^3 ~+~ O(a^4) ~.
\end{eqnarray}  
It is worth commenting on the status of the variables in the renormalization
constants in both schemes. We adopt the convention that the scheme is indicated
on the renormalization constant itself and therefore the coupling constant and
covariant gauge parameter are the variables of the same scheme. However, the 
three loop renormalization of QCD in the RI$^\prime$ scheme has been given in
\cite{32} and the relationship between the parameters has been determined, 
\cite{32}, as  
\begin{equation}   
a_{\mbox{\footnotesize{RI$^\prime$}}} ~=~ 
a_{\mbox{\footnotesize{$\MSbar$}}} ~+~ O \left( 
a_{\mbox{\footnotesize{$\MSbar$}}}^5 \right) 
\end{equation}  
and 
\begin{eqnarray}
\alpha_{\mbox{\footnotesize{RI$^\prime$}}} 
&=& \left[ 1 + \left( \left( - 9 \alpha_{\mbox{\footnotesize{$\MSbar$}}}^2 
- 18 \alpha_{\mbox{\footnotesize{$\MSbar$}}} - 97 \right) C_A + 80 T_F \Nf 
\right) \frac{a_{\mbox{\footnotesize{$\MSbar$}}}}{36} \right. \nonumber \\ 
&& \left. ~+~ \left( \left( 18 \alpha_{\mbox{\footnotesize{$\MSbar$}}}^4 
- 18 \alpha_{\mbox{\footnotesize{$\MSbar$}}}^3 
+ 190 \alpha_{\mbox{\footnotesize{$\MSbar$}}}^2 
- 576 \zeta(3) \alpha_{\mbox{\footnotesize{$\MSbar$}}} 
+ 463 \alpha_{\mbox{\footnotesize{$\MSbar$}}} + 864 \zeta(3) - 7143 \right) 
C_A^2 \right. \right. \nonumber \\ 
&& \left. \left. ~~~~~~~+~ \left( -~ 320 
\alpha_{\mbox{\footnotesize{$\MSbar$}}}^2 
- 320 \alpha_{\mbox{\footnotesize{$\MSbar$}}} + 2304 \zeta(3) 
+ 4248 \right) C_A T_F \Nf \right. \right. \nonumber \\ 
&& \left. \left. ~~~~~~~+~ \left( \frac{}{} -~ 4608 \zeta(3) 
+ 5280 \right) C_F T_F \Nf \right) 
\frac{a^2_{\mbox{\footnotesize{$\MSbar$}}}}{288} \right. \nonumber \\
&& \left. ~+~ \left( \left( ~-~ 486 \alpha_{\mbox{\footnotesize{$\MSbar$}}}^6 
+ 1944 \alpha_{\mbox{\footnotesize{$\MSbar$}}}^5 
+ 4212 \zeta(3) \alpha_{\mbox{\footnotesize{$\MSbar$}}}^4 
- 5670 \zeta(5) \alpha_{\mbox{\footnotesize{$\MSbar$}}}^4 
- 18792 \alpha_{\mbox{\footnotesize{$\MSbar$}}}^4 \right. \right. \right.   
\nonumber \\
&& \left. \left. \left. ~~~~~~~~+~ 48276 \zeta(3) 
\alpha_{\mbox{\footnotesize{$\MSbar$}}}^3 
- 6480 \zeta(5) \alpha_{\mbox{\footnotesize{$\MSbar$}}}^3 
- 75951 \alpha_{\mbox{\footnotesize{$\MSbar$}}}^3 
- 52164 \zeta(3) \alpha_{\mbox{\footnotesize{$\MSbar$}}}^2 
\right. \right. \right. \nonumber \\
&& \left. \left. \left. ~~~~~~~~+~ 2916 \zeta(4) 
\alpha_{\mbox{\footnotesize{$\MSbar$}}}^2 
+ 124740 \zeta(5) \alpha_{\mbox{\footnotesize{$\MSbar$}}}^2 
+ 92505 \alpha_{\mbox{\footnotesize{$\MSbar$}}}^2
- 1303668 \zeta(3) \alpha_{\mbox{\footnotesize{$\MSbar$}}} 
\right. \right. \right. \nonumber \\
&& \left. \left. \left. ~~~~~~~~+~ 11664 \zeta(4) 
\alpha_{\mbox{\footnotesize{$\MSbar$}}} 
+ 447120 \zeta(5) \alpha_{\mbox{\footnotesize{$\MSbar$}}} 
+ 354807 \alpha_{\mbox{\footnotesize{$\MSbar$}}} 
+ 2007504 \zeta(3) 
\right. \right. \right. \nonumber \\
&& \left. \left. \left. ~~~~~~~~+~ 8748 \zeta(4) + 1138050 \zeta(5) 
- 10221367 \right) C_A^3 \right. \right. \nonumber \\
&& \left. \left. ~~~~~~~+~ \left( 
12960 \alpha_{\mbox{\footnotesize{$\MSbar$}}}^4 
- 8640 \alpha_{\mbox{\footnotesize{$\MSbar$}}}^3 
- 129600 \zeta(3) \alpha_{\mbox{\footnotesize{$\MSbar$}}}^2 
- 147288 \alpha_{\mbox{\footnotesize{$\MSbar$}}}^2
\right. \right. \right. \nonumber \\
&& \left. \left. \left. ~~~~~~~~~~~~+~ 698112 \zeta(3) 
\alpha_{\mbox{\footnotesize{$\MSbar$}}} - 312336 
\alpha_{\mbox{\footnotesize{$\MSbar$}}} + 1505088 \zeta(3) - 279936 \zeta(4)
\right. \right. \right. \nonumber \\
&& \left. \left. \left. ~~~~~~~~~~~~-~ 1658880 \zeta(5) + 9236488 \right) 
C_A^2 T_F \Nf  \right. \right. \nonumber \\
&& \left. \left. ~~~~~~~+~ \left( 248832 \zeta(3) 
\alpha_{\mbox{\footnotesize{$\MSbar$}}}^2 
- 285120 \alpha_{\mbox{\footnotesize{$\MSbar$}}}^2 
+ 248832 \zeta(3) \alpha_{\mbox{\footnotesize{$\MSbar$}}} 
- 285120 \alpha_{\mbox{\footnotesize{$\MSbar$}}} 
\right. \right. \right. \nonumber \\
&& \left. \left. \left. ~~~~~~~~~~~~-~ 5156352 \zeta(3) + 373248 \zeta(4)
- 2488320 \zeta(5) + 9293664 \right) C_A C_F T_F \Nf \right. \right. 
\nonumber \\
&& \left. \left. ~~~~~~~+~ \left( ~-~ 38400 
\alpha_{\mbox{\footnotesize{$\MSbar$}}}^2 
- 221184 \zeta(3) \alpha_{\mbox{\footnotesize{$\MSbar$}}} 
+ 55296 \alpha_{\mbox{\footnotesize{$\MSbar$}}}
\right. \right. \right. \nonumber \\
&& \left. \left. \left. ~~~~~~~~~~~~~~\,-~ 884736 \zeta(3) - 1343872 \right) 
C_A T_F^2 \Nf^2 \right. \right. \nonumber \\
&& \left. \left. ~~~~~~~+~ \left( ~-~ 3068928 \zeta(3) + 4976640 \zeta(5) 
- 988416 \right) C_F^2 T_F \Nf \right. \right. \nonumber \\
&& \left. \left. ~~~~~~~+~ \left( 2101248 \zeta(3) - 2842368 \right) 
C_F T_F^2 \Nf^2 \right) 
\frac{a^3_{\mbox{\footnotesize{$\MSbar$}}}}{31104} \right] 
\alpha_{\mbox{\footnotesize{$\MSbar$}}} ~+~ O \left( 
a^4_{\mbox{\footnotesize{$\MSbar$}}} \right) 
\end{eqnarray}
where the scheme of the variable is indicated as a subscript and we note also
that the RI$^\prime$ and $\MSbar$ $\beta$-functions are formally equivalent at 
four loops, \cite{32}.  

From these renormalization constants we can deduce the anomalous dimensions 
from 
\begin{equation}
\gamma^{\mbox{\footnotesize{$\MSbar$}}}_{\bar{\psi}\gamma^\mu D^\nu\psi} 
(a) ~=~ -~ \beta(a) \frac{\partial \ln 
Z^{\mbox{\footnotesize{$\MSbar$}}}_{\bar{\psi}\gamma^\mu D^\nu\psi}} 
{\partial a} ~-~ \alpha \gamma^{\mbox{\footnotesize{$\MSbar$}}}_\alpha(a) 
\frac{\partial \ln 
Z^{\mbox{\footnotesize{$\MSbar$}}}_{\bar{\psi}\gamma^\mu D^\nu\psi}} 
{\partial \alpha} 
\end{equation}
and 
\begin{equation}
\gamma^{\mbox{\footnotesize{RI$^\prime$}}}_{\bar{\psi}\gamma^\mu D^\nu\psi} 
(a) ~=~ -~ \beta(a) \frac{\partial \ln 
Z^{\mbox{\footnotesize{RI$^\prime$}}}_{\bar{\psi}\gamma^\mu D^\nu\psi}} 
{\partial a} ~-~ \alpha \gamma^{\mbox{\footnotesize{RI$^\prime$}}}_\alpha(a) 
\frac{\partial \ln 
Z^{\mbox{\footnotesize{RI$^\prime$}}}_{\bar{\psi}\gamma^\mu D^\nu\psi}} 
{\partial \alpha} ~.  
\end{equation}
We find, in our conventions,  
\begin{eqnarray}  
\gamma^{\mbox{\footnotesize{$\MSbar$}}}_{\bar{\psi} \gamma^\mu D^\nu \psi}(a) 
&=& \frac{8}{3} C_F a ~+~ \frac{8 C_F}{27} \left[ 47 C_A - 14 C_F 
- 16 T_F \Nf \right] a^2 \nonumber \\
&& +~ \frac{8 C_F}{243} \left[ \left( 648 \zeta(3) + 2615 \right) C_A^2 
- \left( 1944 \zeta(3) + 1066 \right) C_A C_F \right. \nonumber \\
&& \left. ~~~~~~~~~~~-~ \left( 1296 \zeta(3) + 782 \right) C_A T_F \Nf 
+ \left( 1296 \zeta(3) - 70 \right) C_F^2 \right. \nonumber \\
&& \left. ~~~~~~~~~~~+~ \left( 1296 \zeta(3) - 853 \right) C_F T_F \Nf 
- 112 T_F^2 \Nf^2 \right] a^3 ~+~ O(a^4) 
\end{eqnarray}  
which agrees with \cite{7} and  
\begin{eqnarray}  
\gamma^{\mbox{\footnotesize{RI$^\prime$}}}_{\bar{\psi} \gamma^\mu D^\nu 
\psi}(a) &=& \frac{8}{3} C_F a ~+~ \frac{C_F}{54} \left[ \left( 27 \alpha^2 
+ 81 \alpha + 1434 \right) C_A - 224 C_F - 504 T_F \Nf \right] a^2 \nonumber \\ 
&& +~ \frac{C_F}{1944} \left[ \left( 486 \alpha^4 + 4131 \alpha^3 
- 972 \zeta(3) \alpha^2 + 19899 \alpha^2 - 10044 \zeta(3) \alpha \right. 
\right. \nonumber \\
&& \left. \left. ~~~~~~~~~~~~+~ 64098 \alpha - 115344 \zeta(3) + 806800 \right)
C_A^2 \right. \nonumber \\
&& \left. ~~~~~~~~~~~-~ \left( 1458 \alpha^2 + 15066 \alpha
+ 10368 \zeta(3) + 172856 \right) C_A C_F \right. \nonumber \\
&& \left. ~~~~~~~~~~~-~ \left( 4320 \alpha^2 - 2592 \zeta(3) \alpha 
+ 22248 \alpha + 25920 \zeta(3) + 534400 \right) C_A T_F \Nf \right. 
\nonumber \\ 
&& \left. ~~~~~~~~~~~+~ \left( 3888 \alpha + 41472 \zeta(3) - 43328 \right) 
C_F T_F \Nf \right. \nonumber \\
&& \left. ~~~~~~~~~~~+~ \left( 82944 \zeta(3) - 4480 \right) C_F^2 
+ 78208 T_F^2 \Nf^2 \right] a^3 ~+~ O(a^4) 
\end{eqnarray}  
in four dimensions. We adopt the same convention for the renormalization group
functions as for the renormalization constants in that the scheme the variables
correspond to is indicated on the left hand side. Restricting to $SU(3)$ and 
the Landau gauge we have 
\begin{eqnarray} 
\left. \gamma^{\mbox{\footnotesize{RI$^\prime$}}}_{\bar{\psi} \gamma^\mu D^\nu 
\psi}(a) \right|_{\alpha = 0}^{SU(3)} &=& \frac{32}{9} a ~-~ \frac{4}{243} 
\left[ 378 \Nf - 6005 \right] a^2 \nonumber \\
&& +~ \frac{8}{6561} \left[ 10998 \Nf^2 - 6318 \zeta(3) \Nf - 467148 \Nf 
\right. \nonumber \\
&& \left. ~~~~~~~~~~~~-~ 524313 \zeta(3) + 3691019 \right] a^3 ~+~ O(a^4) ~. 
\end{eqnarray}  
As an aid to the lattice matching procedure we note that the finite parts of 
the various components in the two schemes after renormalization are 
\begin{eqnarray} 
\left. \Sigma^{(1) ~ {\mbox{\footnotesize{$\MSbar$}}} ~ 
\mbox{\footnotesize{finite}}}_{\bar{\psi} \gamma^\mu 
D^\nu \psi}(p) \right|_{p^2 \, = \, \mu^2} &=& 1 ~-~ \frac{31}{9} C_F a ~-~ 
\left[ \left( \frac{9487}{324} - \frac{5}{3} \alpha - \frac{1}{8} \alpha^2 
- 8 \zeta(3) + 2 \zeta(3) \alpha \right) C_F C_A \right. \nonumber \\
&& \left. ~~~~~~~~~~-~ \frac{2101}{162} T_F \Nf C_F + \left( \frac{1}{2} 
\alpha^2 - \frac{3}{2} \alpha - \frac{8195}{648} + 8 \zeta(3) \right) C_F^2 
\right] a^2 \nonumber \\  
&& -~ \left[ \left( \frac{635}{216} \alpha - \frac{452579}{2187} 
- \frac{976}{81} \zeta(3) - \frac{4}{9} \zeta(3) \alpha - \frac{64}{3} \zeta(4)
\right) T_F \Nf C_F C_A \right. \nonumber \\
&& \left. ~~~~~+~ \left( \frac{85}{9} \alpha - \frac{61322}{2187} 
+ \frac{272}{9} \zeta(3) - \frac{64}{9} \zeta(3) \alpha + \frac{64}{3} \zeta(4)
\right) T_F \Nf C_F^2 \right. \nonumber \\
&& \left. ~~~~~+~ \left( \frac{63602}{2187} + \frac{256}{81} \zeta(3) \right) 
T_F^2 \Nf^2 C_F \right. \nonumber \\
&& \left. ~~~~~+~ \left( \frac{21026833}{69984} - \frac{953}{1728} \alpha 
+ \frac{75}{64} \alpha^2 + \frac{1}{48} \alpha^3 - \left( \frac{135}{4} 
+ \frac{5}{4} \alpha^2 \right) \zeta(5) \right. \right. \nonumber \\
&& \left. \left. ~~~~~~~~~~~+~ \left( \frac{1}{3} \alpha^3 + 2 \alpha^2 
+ \frac{167}{36} \alpha - \frac{3230}{81} \right) \zeta(3) \right. \right. 
\nonumber \\ 
&& \left. \left. ~~~~~~~~~~~+~ \left( \frac{719}{48} - \frac{3}{8} \alpha 
- \frac{3}{16} \alpha^2 \right) \zeta(4) \right) C_F C_A^2 \right. \nonumber \\
&& \left. ~~~~~+~ \left( \frac{5}{8} \alpha^3 - \frac{41}{72} \alpha^2 
- \frac{979}{27} \alpha - \frac{605431}{4374} + \left( \frac{440}{3} 
- \frac{40}{3} \alpha \right) \zeta(5) \right. \right. \nonumber \\
&& \left. \left. ~~~~~~~~~~~-~ \left( \frac{439}{9} - \frac{53}{2} \alpha 
- \frac{1}{6} \alpha^2 + \alpha^3 \right) \zeta(3) - 38 \zeta(4) \right) 
C_F^2 C_A \right. \nonumber \\
&& \left. ~~~~~+~ \left( \left( \frac{7964}{81} - \frac{164}{9} \alpha 
- \frac{16}{3} \alpha^2 + \frac{2}{3} \alpha^3 \right) \zeta(3) + \frac{64}{3} 
\zeta(4) - \frac{400}{3} \zeta(5) \right. \right. \nonumber \\
&& \left. \left. ~~~~~~~~~~~-~ \frac{2935}{5832} + \frac{821}{36} \alpha 
+ \frac{79}{36} \alpha^2 - \frac{1}{4} \alpha^3 \right) C_F^3 \right] a^3 ~+~
O(a^4) 
\end{eqnarray} 
\begin{eqnarray} 
\left. \Sigma^{(2) ~ {\mbox{\footnotesize{$\MSbar$}}} ~ 
\mbox{\footnotesize{finite}}}_{\bar{\psi} \gamma^\mu D^\nu \psi}(p) 
\right|_{p^2 \, = \, \mu^2} &=& -~ \left( \frac{2}{3} + 2 \alpha \right) 
C_F a \nonumber \\
&& -~ \left[ \left( \frac{284}{9} + \frac{29}{3} \alpha + 2 \alpha^2 
- 6 \zeta(3) - 2 \zeta(3) \alpha \right) C_F C_A \right. \nonumber \\
&& \left. ~~~~~-~ \frac{20}{3} T_F \Nf C_F + \left( \alpha^2 - \frac{35}{9} 
\alpha - \frac{74}{3} + 16 \zeta(3) \right) C_F^2 \right] a^2 \nonumber \\  
&& -~ \left[ \left( \frac{184}{3} \zeta(3) + \frac{136}{9} \zeta(3) \alpha
- \frac{2267}{54} \alpha - \frac{87980}{243} \right) T_F \Nf C_F C_A \right. 
\nonumber \\
&& \left. ~~~~~+~ \left( \frac{1399}{81} \alpha + \frac{4913}{27} 
- \frac{1088}{9} \zeta(3) \right) T_F \Nf C_F^2 + \frac{9680}{243} T_F^2 \Nf^2 
C_F \right. \nonumber \\
&& \left. ~~~~~+~ \left( \frac{2762093}{3888} + \frac{28855}{216} \alpha 
+ \frac{431}{16} \alpha^2 + \frac{37}{8} \alpha^3 + \left( \frac{445}{3}
+ 5 \alpha \right) \zeta(5) \right. \right. \nonumber \\
&& \left. \left. ~~~~~~~~~~~-~ \left( \frac{23}{4} \alpha^2 + \frac{1165}{18} 
\alpha + \frac{1617}{4} \right) \zeta(3) \right) C_F C_A^2 \right. \nonumber \\
&& \left. ~~~~~+~ \left( \alpha^3 + \frac{55}{36} \alpha^2 
- \frac{4790}{81} \alpha - \frac{198059}{324} - \left( \frac{1480}{3} 
- \frac{40}{3} \alpha \right) \zeta(5) \right. \right. \nonumber \\
&& \left. \left. ~~~~~~~~~~~+~ \left( \frac{7450}{9} + \frac{131}{9} \alpha 
- 5 \alpha^2 \right) \zeta(3) \right) C_F^2 C_A \right. \nonumber \\
&& \left. ~~~~~+~ \left( \frac{1760}{3} \zeta(5) - \left( \frac{4232}{9} 
+ 8 \alpha \right) \zeta(3) - \frac{29483}{972} \right. \right. \nonumber \\
&& \left. \left. ~~~~~~~~~~~+~ \frac{3281}{324} \alpha + \frac{79}{18} \alpha^2
- \frac{1}{2} \alpha^3 \right) C_F^3 \right] a^3 ~+~ O(a^4) 
\end{eqnarray} 
and 
\begin{eqnarray} 
\left. \Sigma^{(2) ~ {\mbox{\footnotesize{RI$^\prime$}}} ~
\mbox{\footnotesize{finite}} }_{\bar{\psi} 
\gamma^\mu D^\nu \psi}(p) \right|_{p^2 \, = \, \mu^2} &=& -~ \left( \frac{2}{3} 
+ 2 \alpha \right) C_F a \nonumber \\
&& -~ \left[ \left( \frac{284}{9} + \frac{271}{18} \alpha + 3 \alpha^2 
+ \frac{1}{2} \alpha^3 - 6 \zeta(3) - 2 \zeta(3) \alpha \right) C_F C_A \right.
\nonumber \\
&& \left. ~~~~~~-~ \left( \frac{20}{3} + \frac{40}{9} \alpha \right) T_F \Nf 
C_F \right. \nonumber \\
&& \left. ~~~~~~+~ \left( \alpha^2 + 3 \alpha - \frac{604}{27} + 16 \zeta(3) 
\right) C_F^2 \right] a^2 \nonumber \\  
&& -~ \left[ \left( \frac{184}{3} \zeta(3) + \frac{32}{9} \zeta(3) \alpha
- \frac{20}{9} \alpha^3 \right. \right. \nonumber \\
&& \left. \left. ~~~~~~~-~ \frac{40}{3} \alpha^2 - \frac{9470}{81} \alpha 
- \frac{87980}{243} \right) T_F \Nf C_F C_A \right. \nonumber \\
&& \left. ~~~~~-~ \left( \frac{40}{9} \alpha^2 + 52 \alpha - \frac{36536}{243} 
+ \frac{1088}{9} \zeta(3) - 32 \zeta(3) \alpha \right) T_F \Nf C_F^2 \right. 
\nonumber \\
&& \left. ~~~~~+~ \left( \frac{9680}{243} + \frac{800}{81} \alpha \right) 
T_F^2 \Nf^2 C_F \right. \nonumber \\
&& \left. ~~~~~+~ \left( \frac{2762093}{3888} + \frac{289991}{1296} \alpha 
+ \frac{569}{12} \alpha^2 + \frac{127}{9} \alpha^3 + \frac{19}{8} \alpha^4
+ \frac{1}{4} \alpha^5 \right. \right. \nonumber \\
&& \left. \left. ~~~~~~~~~~~-~ \left( \frac{1}{2} \alpha^3 + \frac{11}{4} 
\alpha^2 + \frac{685}{9} \alpha + \frac{1617}{4} \right) \zeta(3) \right. 
\right. \nonumber \\
&& \left. \left. ~~~~~~~~~~~+~ \left( \frac{445}{3} + 5 \alpha \right) \zeta(5)
\right) C_F C_A^2 \right.  \nonumber \\
&& \left. ~~~~~+~ \left( \frac{1}{2} \alpha^4 + \frac{5}{2} \alpha^3 
+ \frac{107}{9} \alpha^2 + \frac{1429}{36} \alpha - \frac{469555}{972} 
\right. \right. \nonumber \\
&& \left. \left. ~~~~~~~~~~~+\, \left( \frac{7216}{9} - 7 \alpha - \alpha^2 
\right) \zeta(3) - \left( \frac{1480}{3} - \frac{40}{3} \alpha 
\right) \zeta(5) \right) C_F^2 C_A \right. \nonumber \\
&& \left. ~~~~~+~ \left( \frac{1760}{3} \zeta(5) - \left( \frac{3688}{9} 
- 8 \alpha \right) \zeta(3) - \frac{56287}{486} \right. \right. \nonumber \\
&& \left. \left. ~~~~~~~~~~~-~ \frac{35}{6} \alpha + \frac{31}{6} \alpha^2 
+ \frac{1}{2} \alpha^3 \right) C_F^3 \right] a^3 ~+~ O(a^4) ~ .  
\end{eqnarray} 
These imply 
\begin{eqnarray} 
\left. \Sigma^{(1) ~ {\mbox{\footnotesize{$\MSbar$}}} ~ 
\mbox{\footnotesize{finite}}}_{\bar{\psi} \gamma^\mu D^\nu \psi}(p) 
\right|^{SU(3) \, , \, \alpha \, = \, 0}_{p^2 \, = \, \mu^2} &=& 1 ~-~ 
\frac{124}{27} a ~-~ \left( \frac{68993}{729} - \frac{160}{9} \zeta(3)
- \frac{2101}{243} \Nf  \right) a^2 \nonumber \\
&& -~ \left[ \frac{451293899}{157464} - \frac{1105768}{2187} \zeta(3)
+ \frac{8959}{324} \zeta(4) + \frac{4955}{81} \zeta(5) \right. \nonumber \\
&& \left. ~~~~~~+~ \left( \frac{224}{81} \zeta(3) - \frac{640}{27} \zeta(4)
- \frac{8636998}{19683} \right) \Nf \right. \nonumber \\
&& \left. ~~~~~~+~ \left( \frac{63602}{6561} + \frac{256}{243} \zeta(3)
\right) \Nf^2 \right] a^3 ~+~ O(a^4) 
\end{eqnarray} 
and 
\begin{eqnarray} 
\left. \Sigma^{(2) ~ {\mbox{\footnotesize{$\MSbar$}}} ~ 
\mbox{\footnotesize{finite}}}_{\bar{\psi} \gamma^\mu D^\nu \psi}(p) 
\right|^{SU(3) \, , \, \alpha \, = \, 0}_{p^2 \, = \, \mu^2} &=& -~ 
\frac{8}{9} a ~-~ \left( \frac{2224}{27} + \frac{40}{9} \zeta(3) 
- \frac{40}{9} \Nf \right) a^2 \nonumber \\
&& -~ \left[ \frac{136281133}{26244} - \frac{376841}{243} \zeta(3)
+ \frac{43700}{81} \zeta(5) \right. \nonumber \\
&& \left. ~~~~~~+~ \left( \frac{1232}{81} \zeta(3) - \frac{15184}{27} \right) 
\Nf + \frac{9680}{729} \Nf^2 \right] a^3 \nonumber \\
&& +~ O(a^4) ~. 
\end{eqnarray} 

\sect{Second moment of transversity operator.} 

The renormalization of the transversity operator $\bar{\psi} \sigma^{\mu\nu} 
D^\rho \psi$ is similar to that for $\bar{\psi} \gamma^\mu D^\nu \psi$ aside
from the decomposition of the corresponding Green's function. The specific 
operator we are interested in is  
\begin{eqnarray} 
{\cal S} \bar{\psi} \sigma^{\mu\nu} D^\rho \psi &=& \bar{\psi} \sigma^{\mu\nu} 
D^\rho \psi ~+~ \bar{\psi} \sigma^{\mu\rho} D^\nu \psi ~-~ \frac{2}{(d-1)} 
\eta^{\nu\rho} \bar{\psi} \sigma^{\mu\lambda} D_\lambda \psi \nonumber \\
&& +~ \frac{1}{(d-1)} \left( \eta^{\mu\nu} \bar{\psi} \sigma^{\rho\lambda} 
D_\lambda \psi + \eta^{\mu\rho} \bar{\psi} \sigma^{\nu\lambda} D_\lambda \psi 
\right) 
\end{eqnarray} 
which is symmetric with respect to the indices $\nu$ and $\rho$ and satisfies 
the traceless conditions, \cite{37},  
\begin{equation} 
{\cal S} \bar{\psi} \sigma^{\mu\nu} D_\mu \psi ~=~ 
{\cal S} \bar{\psi} \sigma^{\mu\nu} D_\nu \psi ~=~ 0 ~.  
\end{equation} 
The Lorentz decomposition of the particular Green's function we will 
renormalize is 
\begin{eqnarray} 
G^{\mu\nu\rho}_{\bar{\psi} \sigma^{\mu\nu} D^\rho \psi}(p) &=&  
\langle \psi(p) ~ [ {\cal S} \bar{\psi} \sigma^{\mu\nu} D^\rho \psi](0) ~ 
\bar{\psi}(-p) \rangle \nonumber \\
&=& \Sigma^{(1)}_{\bar{\psi} \sigma^{\mu\nu} D^\rho \psi}(p) 
\left( \sigma^{\mu\nu} p^\rho + \sigma^{\mu\rho} p^\nu - \frac{(d+2)}{p^2} 
\sigma^{\mu\lambda} p^\nu p^\rho p_\lambda + \eta^{\nu\rho} 
\sigma^{\mu\lambda} p_\lambda \right) \nonumber \\
&& +~ \Sigma^{(2)}_{\bar{\psi} \sigma^{\mu\nu} D^\rho \psi}(p) 
\left( \eta^{\mu\nu} \sigma^{\rho\lambda} p_\lambda + \eta^{\mu\rho} 
\sigma^{\nu\lambda} p_\lambda + \frac{(d-1)(d+2)}{p^2} \sigma^{\mu\lambda} 
p^\nu p^\rho p_\lambda \right. \nonumber \\
&& \left. ~~~~~~~~~~~~~~~~~~~~-~ (d+1) \eta^{\nu\rho} \sigma^{\mu\lambda} 
p_\lambda \right) \nonumber \\ 
&& +~ \Sigma^{(3)}_{\bar{\psi} \sigma^{\mu\nu} D^\rho \psi}(p) 
\left( \sigma^{\nu\lambda} p^\mu p^\rho p_\lambda + \sigma^{\rho\lambda} p^\mu 
p^\nu p_\lambda + d \sigma^{\mu\lambda} p^\nu p^\rho p_\lambda - \eta^{\nu\rho}
\sigma^{\mu\lambda} p_\lambda p^2 \right) \nonumber \\ 
\end{eqnarray} 
where the components are determined through
\begin{eqnarray} 
\Sigma^{(1)}_{\bar{\psi} \sigma^{\mu\nu} D^\rho \psi}(p) &=& 
-~ \frac{1}{8(d-1)(d-2)} \mbox{tr} \left[ \left( \sigma_{\mu\nu} p_\rho
+ \sigma_{\mu\rho} p_\nu - \frac{(d+2)}{p^2} \sigma_{\mu\lambda} p_\nu p_\rho 
p^\lambda \right. \right. \nonumber \\
&& \left. \left. ~~~~~~~~~~~~~~~~~~~~~~~~~~~~+~ \eta_{\nu\rho} 
\sigma_{\mu\lambda} p^\lambda \right) G^{\mu\nu\rho}_{\bar{\psi} 
\sigma^{\mu\nu} D^\rho \psi}(p) \right] \nonumber \\  
&& -~ \frac{1}{8(d-1)(d-2)p^2} \mbox{tr} \left[ \left( \sigma_{\nu\lambda} 
p_\mu p_\rho p^\lambda + \sigma_{\rho\lambda} p_\mu p_\nu p^\lambda \right. 
\right.  \nonumber \\ 
&& \left. \left. ~~~~~~~~~~~~~~~~~~~~~~~~~~~~~~+~ d \sigma_{\mu\lambda} p_\nu 
p_\rho p^\lambda - \eta_{\nu\rho} \sigma_{\mu\lambda} p^\lambda p^2 \right) 
G^{\mu\nu\rho}_{\bar{\psi} \sigma^{\mu\nu} D^\rho \psi}(p) \right] 
\nonumber \\  
\Sigma^{(2)}_{\bar{\psi} \sigma^{\mu\nu} D^\rho \psi}(p) &=& 
-~ \frac{(d^2-d+2)}{8(d-1)(d^2-1)(d^2-4)} \mbox{tr} \left[ \left( 
\eta_{\mu\nu} \sigma_{\rho\lambda} p^\lambda + \eta_{\mu\rho} 
\sigma_{\nu\lambda} p^\lambda - (d+1) \eta_{\nu\rho} \sigma_{\mu\lambda} 
p^\lambda \right. \right. \nonumber \\
&& \left. \left. ~~~~~~~~~~~~~~~~~~~~~~~~~~~~~~~~~~~~~~~~~+ 
\frac{(d-1)(d+2)}{p^2} \sigma_{\mu\lambda} p_\nu p_\rho p^\lambda \right) 
G^{\mu\nu\rho}_{\bar{\psi} \sigma^{\mu\nu} D^\rho \psi}(p) \right] \nonumber \\
&& +~ \frac{1}{8(d+1)(d-2)p^2} \mbox{tr} \left[ \left( \sigma_{\nu\lambda} 
p_\mu p_\rho p^\lambda + \sigma_{\rho\lambda} p_\mu p_\nu p^\lambda \right. 
\right. \nonumber \\ 
&& \left. \left. ~~~~~~~~~~~~~~~~~~~~~~~~~~~~~~+~ d \sigma_{\mu\lambda} p_\nu 
p_\rho p^\lambda - \eta_{\nu\rho} \sigma_{\mu\lambda} p^\lambda p^2 \right) 
G^{\mu\nu\rho}_{\bar{\psi} \sigma^{\mu\nu} D^\rho \psi}(p) \right] \nonumber \\ 
\Sigma^{(3)}_{\bar{\psi} \sigma^{\mu\nu} D^\rho \psi}(p) &=& 
-~ \frac{1}{8(d-1)(d-2)} \mbox{tr} \left[ \left( \sigma_{\mu\nu} p_\rho
+ \sigma_{\mu\rho} p_\nu - \frac{(d+2)}{p^2} \sigma_{\mu\lambda} p_\nu p_\rho 
p^\lambda \right. \right. \nonumber \\
&& \left. \left. ~~~~~~~~~~~~~~~~~~~~~~~~~~~~+~ \eta_{\nu\rho} 
\sigma_{\mu\lambda} p^\lambda \right) G^{\mu\nu\rho}_{\bar{\psi} 
\sigma^{\mu\nu} D^\rho \psi}(p) \right] \nonumber \\  
&& +~ \frac{1}{8(d+1)(d-2)} \mbox{tr} \left[ \left( 
\eta_{\mu\nu} \sigma_{\rho\lambda} p^\lambda + \eta_{\mu\rho} 
\sigma_{\nu\lambda} p^\lambda - (d+1) \eta_{\nu\rho} \sigma_{\mu\lambda} 
p^\lambda \right. \right. \nonumber \\
&& \left. \left. ~~~~~~~~~~~~~~~~~~~~~~~~~~~~+~ \frac{(d-1)(d+2)}{p^2} 
\sigma_{\mu\lambda} p_\nu p_\rho p^\lambda \right) G^{\mu\nu\rho}_{\bar{\psi} 
\sigma^{\mu\nu} D^\rho \psi}(p) \right] \nonumber \\  
&& -~ \frac{(d+2)}{8(d+1)(d-2)p^2} \mbox{tr} \left[ \left( \sigma_{\nu\lambda} 
p_\mu p_\rho p^\lambda + \sigma_{\rho\lambda} p_\mu p_\nu p^\lambda \right. 
\right.  \nonumber \\ 
&& \left. \left. ~~~~~~~~~~~~~~~~~~~~~~~~~~~~~~+~ d \sigma_{\mu\lambda} p_\nu 
p_\rho p^\lambda - \eta_{\nu\rho} \sigma_{\mu\lambda} p^\lambda p^2 \right) 
G^{\mu\nu\rho}_{\bar{\psi} \sigma^{\mu\nu} D^\rho \psi}(p) \right] ~. 
\end{eqnarray} 
Using the same programmes which determined the renormalization of $\bar{\psi} 
\gamma^\mu D^\nu \psi$ but instead including the Feynman rule for the 
transversity operator, we find after renormalizing the Green's function in the 
$\MSbar$ scheme that  
\begin{eqnarray} 
Z^{\mbox{\footnotesize{$\MSbar$}}}_{\bar{\psi} \sigma^{\mu\nu} D^\rho \psi} 
&=& 1 ~+~ \frac{3}{\epsilon} C_F a \nonumber \\
&& +~ \left[ \left( 4 T_F \Nf + 9 C_F - 11 C_A \right) 
\frac{1}{2\epsilon^2} ~+~ \left( 35 C_A - 9 C_F - 12 T_F \Nf \right)
\frac{1}{4\epsilon} \right] C_F a^2 \nonumber \\
&& +~ \left[ \! \left( 242 C_A^2 - 297 C_A C_F - 176 C_A T_F \Nf + 81 C_F^2 
+ 108 C_F T_F \Nf + 32 T_F^2 \Nf^2 \right) \frac{1}{18\epsilon^3} \right. 
\nonumber \\
&& \left. ~~~~+~ \left( 1143 C_A C_F - 1178 C_A^2 + 784 C_A T_F \Nf \right.
\right. \nonumber \\
&& \left. \left. ~~~~~~~~~~-~ 243 C_F^2 - 252 C_F T_F \Nf - 96 T_F^2 \Nf^2
\right) \frac{1}{36\epsilon^2} \right. \nonumber \\  
&& \left. ~~~~+~ \left( 12553 C_A^2 - 7479 C_A C_F - \left( 5184 \zeta(3) 
+ 4168 \right) C_A T_F \Nf + 1782 C_F^2 \right. \right. \nonumber \\
&& \left. \left. ~~~~~~~~~~+~ \left( 5184 \zeta(3) - 3240 \right) C_F T_F 
\Nf - 368 T_F^2 \Nf^2 \right) \frac{1}{324\epsilon} \right] C_F a^3 ~+~ 
O(a^4) \nonumber \\  
\end{eqnarray} 
and with the RI$^\prime$ scheme definition  
\begin{equation}  
\left. \lim_{\epsilon \, \rightarrow \, 0} \left[ 
Z^{\mbox{\footnotesize{RI$^\prime$}}}_\psi  
Z^{\mbox{\footnotesize{RI$^\prime$}}}_{\bar{\psi} \sigma^{\mu\nu} D^\rho \psi}  
\Sigma^{(1)}_{\bar{\psi} \sigma^{\mu\nu} D^\rho \psi}(p) \right] 
\right|_{p^2 \, = \, \mu^2} ~=~ 1 
\end{equation}  
we have 
\begin{eqnarray} 
Z^{\mbox{\footnotesize{RI$^\prime$}}}_{\bar{\psi} \sigma^{\mu\nu} D^\rho \psi} 
&=& 1 ~+~ \left[ \frac{3}{\epsilon} + \frac{1}{2} \left( 3 \alpha + 7 \right)
\right] C_F a \nonumber \\
&& +~ \left[ \left( 4 T_F \Nf + 9 C_F - 11 C_A \right) 
\frac{1}{2\epsilon^2} ~+~ \left( 35 C_A + ( 33 + 18 \alpha ) C_F - 12 T_F \Nf 
\right) \frac{1}{4\epsilon} \right. \nonumber \\
&& \left. ~~~~~+~ \frac{1}{24} \left( \left( 9 \alpha^3 + 54 \alpha^2 
- 36 \zeta(3) \alpha + 271 \alpha - 324 \zeta(3) + 1189 \right) C_A 
\right. \right. \nonumber \\
&& \left. \left. ~~~~~~~~~~~~~~~~~+ \left( 36 \alpha^2 + 108 \alpha 
+ 288 \zeta(3) - 165 \right) C_F - \left( 80 \alpha + 468 \right) T_F \Nf 
\right) \right] a^2 \nonumber \\ 
&& +~ \left[ \! \left( 242 C_A^2 - 297 C_A C_F - 176 C_A T_F \Nf 
+ 81 C_F^2 + 108 C_F T_F \Nf + 32 T_F^2 \Nf^2 \right) \frac{1}{18\epsilon^3}
\right. \nonumber \\
&& \left. ~~~~+~ \left( ( 450 - 297 \alpha ) C_A C_F - 1178 C_A^2 + 784 C_A T_F
\Nf + ( 243 \alpha + 324 ) C_F^2 \right. \right. \nonumber \\
&& \left. \left. ~~~~~~~~~~~~~~~+~ 108 \alpha C_F T_F \Nf - 96 T_F^2 \Nf^2 
\right) \frac{1}{36\epsilon^2} \right. \nonumber \\  
&& \left. ~~~~+~ \left( \left( 729 \alpha^3 + 4374 \alpha^2 
- 2916 \zeta(3) \alpha + 30456 \alpha - 26244 \zeta(3) + 101196 \right) C_A C_F 
\right. \right. \nonumber \\
&& \left. \left. ~~~~~~~~~~~~~~~+~ 25106 C_A^2 - \left( 10368 \zeta(3) 
+ 8336 \right) C_A T_F \Nf \right. \right. \nonumber \\
&& \left. \left. ~~~~~~~~~~~~~~~+~ \left( 2916 \alpha^2 + 6561 \alpha - 14904
+ 23328 \zeta(3) \right) C_F^2 - 736 T_F^2 \Nf^2 \right. \right. \nonumber \\
&& \left. \left. ~~~~~~~~~~~~~~~+~ \left( 10368 \zeta(3) - 51192 - 9396 \alpha 
\right) C_F T_F \Nf \right) \frac{1}{648\epsilon} \right. \nonumber \\
&& \left. ~~~~+~ \frac{1}{5184} \left( \left( 972 \alpha^5 + 9234 \alpha^4 
- 1944 \zeta(3) \alpha^3 + 54864 \alpha^3 - 10692 \zeta(3) \alpha^2
\right. \right. \right. \nonumber \\ 
&& \left. \left. \left. ~~~~~~~~~~~~~~~~~~~+~ 184356 \alpha^2 
- 282312 \zeta(3) \alpha + 19440 \zeta(5) \alpha + 874185 \alpha \right. 
\right. \right. \nonumber \\
&& \left. \left. \left. ~~~~~~~~~~~~~~~~~~~-~ 1644228 \zeta(3) 
+ 447120 \zeta(5) + 3930425 \right) C_A^2 \right. \right. \nonumber \\
&& \left. \left. ~~~~~~~~~~~~~~~~~~~+~ \left( 3888 \alpha^4 + 24300 \alpha^3 
+ 3888 \zeta(3) \alpha^2 + 116748 \alpha^2 \right. \right. \right. 
\nonumber \\
&& \left. \left. \left. ~~~~~~~~~~~~~~~~~~~~~~~~-~ 34992 \zeta(3) \alpha 
+ 51840 \zeta(5) \alpha + 375732 \alpha + 1260576 \zeta(3) \right. \right. 
\right. \nonumber \\
&& \left. \left. \left. ~~~~~~~~~~~~~~~~~~~~~~~~-~ 466560 \zeta(5) - 511188 
\right) C_A C_F \right. \right. \nonumber \\
&& \left. \left. ~~~~~~~~~~~~~~~~~~~-~ \left( 8640 \alpha^3 + 51840 \alpha^2 
- 13824 \zeta(3) \alpha + 454560 \alpha \right. \right. \right. \nonumber \\
&& \left. \left. \left. ~~~~~~~~~~~~~~~~~~~~~~~~-~ 124992 \zeta(3) 
+ 124416 \zeta(4) + 2438256 \right) C_A T_F \Nf \right. \right. \nonumber \\
&& \left. \left. ~~~~~~~~~~~~~~~~~~~+~ \left( 38400 \alpha + 18432 \zeta(3) 
+ 323200 \right) T_F^2 \Nf^2 \right. \right. \nonumber \\ 
&& \left. \left. ~~~~~~~~~~~~~~~~~~~-~ \left( 34560 \alpha^2 - 82944 \zeta(3)
\alpha + 295920 \alpha - 82944 \zeta(3) \right. \right. \right. \nonumber \\
&& \left. \left. \left. ~~~~~~~~~~~~~~~~~~~~~~~~-~ 124416 \zeta(4) 
+ 620592 \right) C_F T_F \Nf \right. \right. \nonumber \\
&& \left. \left. ~~~~~~~~~~~~~~~~~~~+~ \left( 7776 \alpha^3 - 31104\zeta(3) 
\alpha^2 + 54432 \alpha^2 + 32076 \alpha + 51840 \zeta(3) \right. \right. 
\right. \nonumber \\ 
&& \left. \left. \left. ~~~~~~~~~~~~~~~~~~~~~~~~+~ 207360 \zeta(5) - 266868 
\right) C_F^2 \right) \right] C_F a^3 ~+~ O(a^4) ~.  
\end{eqnarray}  
Consequently, the renormalization group function 
\begin{eqnarray}  
\gamma^{\mbox{\footnotesize{$\MSbar$}}}_{\bar{\psi} \sigma^{\mu\nu} D^\rho 
\psi}(a) &=& 3 C_F a ~+~ \frac{C_F}{2} \left[ 35 C_A - 9 C_F - 12 T_F \Nf 
\right] a^2 \nonumber \\
&& +~ \frac{C_F}{108} \left[ 12553 C_A^2 - 7479 C_A C_F - \left( 5184 \zeta(3) 
+ 4168 \right) C_A T_F \Nf \right. \nonumber \\
&& \left. ~~~~~~~~~+~ 1782 C_F^2 + \left( 5184 \zeta(3) - 3240 \right) 
C_F T_F \Nf - 368 T_F^2 \Nf^2 \right] a^3 \nonumber \\ 
&& +~ O(a^4) 
\label{gamtramsb} 
\end{eqnarray}  
has emerged as being independent of $\alpha$ as expected. The two loop part of 
(\ref{gamtramsb}) agrees with the known results, \cite{38,39,40,37,41}, when 
they are restricted to the second moment. A second check on the result is that
the double and triple poles in $\epsilon$ at three loops in the renormalization
constant have correctly emerged. Also, in four dimensions  
\begin{eqnarray}  
\gamma^{\mbox{\footnotesize{RI$^\prime$}}}_{\bar{\psi} \sigma^{\mu\nu} D^\rho
\psi}(a) &=& 3 C_F a ~+~ \frac{C_F}{12} \left[ \left( 9 \alpha^2 + 27 \alpha 
+ 364 \right) C_A - 54 C_F - 128 T_F \Nf \right] a^2 \nonumber \\ 
&& +~ \frac{C_F}{432} \left[ \left( 162 \alpha^4 + 1377 \alpha^3 
- 324 \zeta(3) \alpha^2 + 6633 \alpha^2 - 3348 \zeta(3) \alpha \right. 
\right. \nonumber \\
&& \left. \left. ~~~~~~~~~~~~+~ 21366 \alpha - 42768 \zeta(3) + 224296 \right)
C_A^2 \right. \nonumber \\
&& \left. ~~~~~~~~~~~+~ \left( 162 \alpha^3 + 324 \alpha^2 - 1674 \alpha
+ 38016 \zeta(3) - 71100 \right) C_A C_F \right. \nonumber \\
&& \left. ~~~~~~~~~~~-~ \left( 1440 \alpha^2 - 864 \zeta(3) \alpha 
+ 7416 \alpha + 5184 \zeta(3) + 145600 \right) C_A T_F \Nf \right. 
\nonumber \\ 
&& \left. ~~~~~~~~~~~+~ \left( 432 \alpha + 6912 \zeta(3) - 4032 \right) 
C_F T_F \Nf \right. \nonumber \\
&& \left. ~~~~~~~~~~~+~ 7128 C_F^2 + 20992 T_F^2 \Nf^2 \right] a^3 ~+~ 
O(a^4) ~.  
\end{eqnarray}  
Thus, for $SU(3)$ and the Landau gauge we have 
\begin{eqnarray} 
\left. \gamma^{\mbox{\footnotesize{$\MSbar$}}}_{\bar{\psi} \sigma^{\mu\nu} 
D^\rho \psi}(a) \right|_{\alpha = 0}^{SU(3)} &=& 4 a ~-~ 2 \left[ 2 \Nf - 31 
\right] a^2 \nonumber \\
&& -~ \frac{1}{81} \left[ 92 \Nf^2 + 4320 \zeta(3) \Nf + 8412 \Nf - 86229 
\right] a^3 ~+~ O(a^4)  
\end{eqnarray}  
and 
\begin{eqnarray} 
\left. \gamma^{\mbox{\footnotesize{RI$^\prime$}}}_{\bar{\psi} \sigma^{\mu\nu} 
D^\rho \psi}(a) \right|_{\alpha = 0}^{SU(3)} &=& 4 a ~-~ \frac{4}{9} \left[ 
16 \Nf - 255 \right] a^2 \nonumber \\
&& +~ \frac{2}{81} \left[ 656 \Nf^2 - \left( 396 \zeta(3) + 27636 \right) 
\Nf - 29106 \zeta(3) + 218367 \right] a^3 \nonumber \\ 
&& +~ O(a^4) 
\end{eqnarray}  
which are the main results of this article. To assist with the lattice matching
procedure we record that the finite parts of the components in both the 
$\MSbar$ and RI$^\prime$ schemes after renormalization are 
\begin{eqnarray} 
\left. \Sigma^{(1) ~ {\mbox{\footnotesize{$\MSbar$}}} ~ 
\mbox{\footnotesize{finite}}}_{\bar{\psi} \sigma^{\mu\nu} D^\rho \psi}(p) 
\right|_{p^2 \, = \, \mu^2} &=& 1 ~-~ \left( \frac{7}{2} + \frac{\alpha}{2} 
\right) C_F a \nonumber \\
&& -~ \left[ \left( \frac{943}{24} + \frac{3}{4} \alpha + \frac{3}{8} \alpha^2
- \frac{21}{2} \zeta(3) + \frac{3}{2} \zeta(3) \alpha \right) C_F C_A 
- 16 T_F \Nf C_F \right. \nonumber \\
&& \left. ~~~~~-~ \left( \frac{37}{2} + \frac{5}{2} \alpha - \frac{3}{4} 
\alpha^2 - 12 \zeta(3) \right) C_F^2 \right] a^2 \nonumber \\   
&& -~ \left[ \left( \frac{478}{27} + \frac{47}{3} \alpha - 8 \zeta(3) \alpha
+ 24 \zeta(4) \right) T_F \Nf C_F^2 \right. \nonumber \\ 
&& \left. ~~~~~+~ \left( \frac{1160}{27} + \frac{32}{9} \zeta(3) \right) 
T_F^2 \Nf^2 C_F \right. \nonumber \\
&& \left. ~~~~~+~ \left( \! \left( \frac{61}{9} + \frac{10}{3} \alpha \right) 
\! \zeta(3) - 24 \zeta(4) - \frac{104843}{324} - \frac{68}{9} \! \alpha \right) 
T_F \Nf C_F C_A \right. \nonumber \\
&& \left. ~~~~~+~ \left( \frac{2656369}{5184} + \frac{18379}{576} \alpha 
+ \frac{253}{32} \alpha^2 + \frac{113}{96} \alpha^3 \right. \right. 
\nonumber \\
&& \left. \left. ~~~~~~~~~~~-~ \left( \frac{26839}{144} + \frac{131}{12} \alpha 
- \frac{9}{16} \alpha^2 - \frac{1}{3} \alpha^3 \right) \zeta(3) \right. \right.
\nonumber \\
&& \left. \left. ~~~~~~~~~~~+~ \left( \frac{69}{16} - \frac{3}{8} \alpha 
- \frac{3}{16} \alpha^2 \right) \zeta(4) \right. \right. \nonumber \\
&& \left. \left. ~~~~~~~~~~~+~ \left( 45 + \frac{5}{4} \alpha - \frac{5}{4} 
\alpha^2 \right) \zeta(5) \right) C_F C_A^2 \right. \nonumber \\
&& \left. ~~~~~+~ \left( \left( \frac{1699}{6} + \frac{131}{4} \alpha 
- \frac{3}{4} \alpha^2 - \alpha^3 \right) \zeta(3)  
- \left( 70 + 10 \alpha \right) \zeta(5) \right. \right. \nonumber \\
&& \left. \left. ~~~~~~~~~~~-~ 6 \zeta(4) - \frac{158969}{432} 
- \frac{2657}{48} \alpha - \frac{5}{8} \alpha^2 + \frac{7}{8} \alpha^3 \right) 
C_F^2 C_A \right. \nonumber \\
&& \left. ~~~~~+~ \left( 40 \zeta(5) - \left( 74 + 24 \alpha + 6 \alpha^2 
- \frac{2}{3} \alpha^3 \right) \zeta(3) \right. \right. \nonumber \\
&& \left. \left. ~~~~~~~~~~~+~ \frac{521}{12} + \frac{59}{2} \alpha 
+ \frac{33}{8} \alpha^2 - \frac{3}{8} \alpha^3 \right) C_F^3 \right] a^3 ~+~ 
O(a^4)
\end{eqnarray} 
\begin{equation} 
\left. \Sigma^{(2) ~ {\mbox{\footnotesize{$\MSbar$}}} ~ 
\mbox{\footnotesize{finite}}}_{\bar{\psi} \sigma^{\mu\nu} D^\rho \psi}(p) 
\right|_{p^2 \, = \, \mu^2} ~=~ \frac{1}{3}  
\left. \Sigma^{(1) ~ {\mbox{\footnotesize{$\MSbar$}}} ~ 
\mbox{\footnotesize{finite}}}_{\bar{\psi} \sigma^{\mu\nu} D^\rho \psi}(p) 
\right|_{p^2 \, = \, \mu^2} ~+~ O(a^4)  
\end{equation} 
and 
\begin{equation} 
\left. \Sigma^{(3) ~ {\mbox{\footnotesize{$\MSbar$}}} ~ 
\mbox{\footnotesize{finite}}}_{\bar{\psi} \sigma^{\mu\nu} D^\rho \psi}(p) 
\right|_{p^2 \, = \, \mu^2} ~=~  
\left. \Sigma^{(3) ~ {\mbox{\footnotesize{RI$^\prime$}}} ~ 
\mbox{\footnotesize{finite}}}_{\bar{\psi} \sigma^{\mu\nu} D^\rho \psi}(p) 
\right|_{p^2 \, = \, \mu^2} ~=~ O(a^4) 
\end{equation} 
to three loops. Hence,  
\begin{eqnarray} 
\left. \Sigma^{(1) ~ {\mbox{\footnotesize{$\MSbar$}}} ~ 
\mbox{\footnotesize{finite}}}_{\bar{\psi} \sigma^{\mu\nu} D^\rho \psi}(p) 
\right|^{SU(3) \, , \, \alpha \, = \, 0}_{p^2 \, = \, \mu^2} &=&  1 ~-~
\frac{14}{3} a ~-~ \left( \frac{2237}{18} - \frac{62}{3} \zeta(3)
- \frac{32}{3} \Nf \right) a^2 \nonumber \\
&& -~ \left[ \frac{1852993}{432} - \frac{97391}{108} \zeta(3)
+ \frac{79}{4} \zeta(4) + \frac{7060}{27} \zeta(5) \right. \nonumber \\
&& \left. ~~~~~~+~ \left( \frac{122}{9} \zeta(3) - \frac{80}{3} \zeta(4) 
- \frac{306881}{486} \right) \Nf \right. \nonumber \\
&& \left. ~~~~~~+~ \left( \frac{1160}{81} + \frac{32}{27} \zeta(3) 
\right) \Nf^2 \right] a^3 ~+~ O(a^4) ~. 
\end{eqnarray} 

\sect{Conversion functions.} 

As in \cite{32} we have checked our expressions by also constructing the 
conversion functions $C_{\cal O}(a,\alpha)$ for each of the operators
explicitly. These functions allow one to move from one scheme to another and 
are defined by, \cite{42},  
\begin{equation} 
C_{\cal O}(a,\alpha) ~=~ 
\frac{Z^{\mbox{\footnotesize{RI$^\prime$}}}_{\cal O}} 
{Z^{\mbox{\footnotesize{$\MSbar$}}}_{\cal O}} ~. 
\end{equation} 
The renormalization group functions are then related by 
\begin{eqnarray}
\gamma^{\mbox{\footnotesize{RI$^\prime$}}}_{\cal O} 
\left(a_{\mbox{\footnotesize{RI$^\prime$}}}\right) &=& 
\gamma^{\mbox{\footnotesize{$\MSbar$}}}_{\cal O} 
\left(a_{\mbox{\footnotesize{$\MSbar$}}}\right) ~-~ 
\beta\left(a_{\mbox{\footnotesize{$\MSbar$}}}\right) 
\frac{\partial ~}{\partial a_{\mbox{\footnotesize{$\MSbar$}}}} 
\ln C_{\cal O} \left(a_{\mbox{\footnotesize{$\MSbar$}}}, 
\alpha_{\mbox{\footnotesize{$\MSbar$}}}\right) \nonumber \\
&& -~ \alpha_{\mbox{\footnotesize{$\MSbar$}}} 
\gamma^{\mbox{\footnotesize{$\MSbar$}}}_\alpha 
\left(a_{\mbox{\footnotesize{$\MSbar$}}}\right) 
\frac{\partial ~}{\partial \alpha_{\mbox{\footnotesize{$\MSbar$}}}}  
\ln C_{\cal O} \left(a_{\mbox{\footnotesize{$\MSbar$}}},  
\alpha_{\mbox{\footnotesize{$\MSbar$}}}\right) 
\end{eqnarray} 
where we have indicated the scheme of the variables explicitly. Being careful 
to first convert the variables to the {\em same} reference scheme, which we 
will take to be the $\MSbar$ scheme, we have from the various renormalization 
constants, 
\begin{eqnarray} 
C_{\bar{\psi} \gamma^\mu D^\nu \psi}(a,\alpha) &=& 1 ~+~ \left( 9 \alpha + 31
\right) \frac{C_F a}{9} \nonumber \\
&& +~ \left[ \left( 162 \alpha^2 - 162 \zeta(3) \alpha + 783 \alpha - 1782 
\zeta(3) + 6404 \right) C_A \right. \nonumber \\
&& \left. ~~~~+~ \left( 81 \alpha^2 + 315 \alpha + 1296 \zeta(3) - 228 \right)
C_F - 2668 T_F \Nf \right] \frac{C_F a^2}{162} \nonumber \\
&& +~ \left[ \left( 161838 \alpha^3 - 201204 \zeta(3) \alpha^2  
+ 942597 \alpha^2 - 2124792 \zeta(3) \alpha \right. \right. \nonumber \\
&& \left. \left. ~~~~~~~+~ 174960 \zeta(5) \alpha + 4796982 \alpha 
- 11944044 \zeta(3) + 746496 \zeta(4) \right. \right. \nonumber \\
&& \left. \left. ~~~~~~~+~ 524880 \zeta(5) + 38226589 \right) C_A^2 \right. 
\nonumber \\
&& \left. ~~~~+~ \left( 26244 \alpha^3 + 151632 \zeta(3) \alpha^2 
+ 159408 \alpha^2 + 346032 \zeta(3) \alpha \right. \right. \nonumber \\
&& \left. \left. ~~~~~~~~~~+~ 466560 \zeta(5) \alpha + 554904 \alpha 
- 4914432 \zeta(3) \right. \right. \nonumber \\
&& \left. \left. ~~~~~~~~~~-~ 2239488 \zeta(4) + 8864640 \zeta(5) 
+ 3993332 \right) C_A C_F \right. \nonumber \\
&& \left. ~~~~+~ \left( 528768 \zeta(3) \alpha - 1469016 \alpha 
+ 369792 \zeta(3) \right. \right. \nonumber \\
&& \left. \left. ~~~~~~~~~~-~ 1492992 \zeta(4) - 24752896 \right) C_A T_F \Nf 
\right. \nonumber \\
&& \left. ~~~~+~ \left( 17496 \alpha^3 - 373248 \zeta(3) \alpha^2 
+ 289656 \alpha^2 - 715392 \zeta(3) \alpha \right. \right. \nonumber \\
&& \left. \left. ~~~~~~~~~~+~ 879336 \alpha + 10737792 \zeta(3) 
+ 1492992 \zeta(4) \right. \right. \nonumber \\
&& \left. \left. ~~~~~~~~~~-~ 9331200 \zeta(5) - 3848760 \right) C_F^2 \right. 
\nonumber \\
&& \left. ~~~~-~ \left( 497664 \zeta(3) \alpha + 351648 \alpha  
- 3234816 \zeta(3) \right. \right. \nonumber \\
&& \left. \left. ~~~~~~~~~~-~ 1492992 \zeta(4) + 9980032 \right) C_F T_F \Nf 
\right. \nonumber \\
&& \left. ~~~~+~ \left( 221184 \zeta(3) + 3391744 \right) T_F^2 \Nf^2 \right] 
\frac{C_F a^3}{69984} ~+~ O(a^4)  
\end{eqnarray} 
and 
\begin{eqnarray} 
C_{\bar{\psi} \sigma^{\mu\nu} D^\rho \psi}(a,\alpha) &=& 1 ~+~ \left( 3 \alpha 
+ 7 \right) \frac{C_F a}{2} \nonumber \\
&& +~ \left[ \left( 36 \alpha^2 - 36 \zeta(3) \alpha + 174 \alpha 
- 324 \zeta(3) + 1189 \right) C_A \right. \nonumber \\
&& \left. ~~~~+~ \left( 36 \alpha^2 + 108 \alpha + 288 \zeta(3) - 165 \right)
C_F - 486 T_F \Nf \right] \frac{C_F a^2}{24} 
\nonumber \\
&& +~ \left[ \left( 17982 \alpha^3 - 22356 \zeta(3) \alpha^2 
+ 104733 \alpha^2 - 238032 \zeta(3) \alpha + 19440 \zeta(5) \alpha \right.
\right. \nonumber \\
&& \left. \left. ~~~~~~~~~~+~ 523602 \alpha - 1644228 \zeta(3) 
+ 447120 \zeta(5) + 3930425 \right) C_A^2 \right. \nonumber \\
&& \left. ~~~~+~ \left( 10692 \alpha^3 + 3888 \zeta(3) \alpha^2 
+ 63180 \alpha^2 - 34992 \zeta(3) \alpha \right. \right. \nonumber \\
&& \left. \left. ~~~~~~~~~~+~ 51840 \zeta(5) \alpha + 312876 \alpha 
+ 1260576 \zeta(3) \right. \right. \nonumber \\
&& \left. \left. ~~~~~~~~~~-~ 466560 \zeta(5) - 511188 \right) C_A C_F 
\right. \nonumber \\
&& \left. ~~~~+~ \left( 58752 \zeta(3) \alpha - 163224 \alpha + 124992 \zeta(3)
\right. \right. \nonumber \\
&& \left. \left. ~~~~~~~~~~-~ 124416 \zeta(4) - 2438256 \right) C_A T_F \Nf 
\right. \nonumber \\
&& \left. ~~~~+~ \left( 7776 \alpha^3 - 31104 \zeta(3) \alpha^2 
+ 54432 \alpha^2 + 32076 \alpha \right. \right. \nonumber \\
&& \left. \left. ~~~~~~~~~~+~ 51840 \zeta(3) + 207360 \zeta(5) - 266868 \right)
C_F^2 \right. \nonumber \\
&& \left. ~~~~-~ \left( 41472 \zeta(3) \alpha + 101520 \alpha - 82944 \zeta(3) 
\right. \right. \nonumber \\
&& \left. \left. ~~~~~~~~~~-~ 124416 \zeta(4) + 620592 \right) C_F T_F \Nf 
\right. \nonumber \\
&& \left. ~~~~+~ \left( 18432 \zeta(3) + 323200 \right) T_F^2 \Nf^2 \right] 
\frac{C_F a^3}{5184} ~+~ O(a^4) ~. 
\end{eqnarray} 
With these expressions given in terms of the $\MSbar$ coupling constant and
covariant gauge parameter, we have checked that the same RI$^\prime$ scheme
anomalous dimensions emerge for each operator and for non-zero $\alpha$. 

\sect{Large $\Nf$ critical exponent for transversity operator.} 

In order to provide another independent check on our transversity results for
$\MSbar$ we have extended the leading order large $\Nf$ calculation of the 
critical exponent associated with the flavour non-singlet twist-$2$ operator 
anomalous dimension as a function of $d$ $=$ $2\mu$ and {\em arbitrary} moment 
$n$, \cite{43}, to the transversity case. In other words we consider the 
operator $\bar{\psi} \sigma^{\mu\nu_1} D^{\nu_2} \ldots D^{\nu_{n-1}} \psi$ 
which is symmetric in the indices $\{\nu_i\}$ and satisfies a more general 
traceless condition to that for $n$~$=$~$2$, \cite{37}. As the large $\Nf$ 
critical point technique has been widely documented in \cite{44,45,43} we refer
interested readers to these sources and quote the main result of our
computations. Following the procedure of \cite{43}, we found the 
$d$-dimensional critical exponent for the transversity operator is  
\begin{eqnarray} 
\eta^{(n)}_{\bar{\psi} \sigma^{\mu\nu} D^\rho \psi}
&=& \frac{2(\mu-1)C_F \eta^{\mbox{\footnotesize{o}}}_1}{(2\mu-1)(\mu-2) 
T_F \Nf} \left[ (\mu-1) ~-~ \frac{\mu(\mu-2)^2}{(\mu+n-1)(\mu+n-2)} \right. 
\nonumber \\
&& \left. ~~~~~~~~~~~~~~~~~~~~~~~~~~~~~+~ 2 \mu \left[ \psi(\mu+n-1) 
- \psi(\mu) \right] \frac{}{} \right] ~+~ O \left( \frac{1}{\Nf^2} \right) 
\label{traexp} 
\end{eqnarray} 
where 
\begin{equation} 
\eta^{\mbox{\footnotesize{o}}}_1 ~=~ \frac{(2\mu-1)(\mu-2)\Gamma(2\mu)}
{4\Gamma^2(\mu)\Gamma(\mu+1)\Gamma(2-\mu)} 
\end{equation} 
and $\psi(x)$ $=$ $d \ln \Gamma(x)/dx$ and $\Gamma(x)$ is the Euler gamma
function. If one expands (\ref{traexp}) in powers of $\epsilon$, $\mu$ $=$ $2$
$-$ $\epsilon$, then there is a one-to-one correspondence with the leading
large $\Nf$ coefficients of the associated anomalous dimension to all orders in
the (perturbative) coupling constant. Moreover, since we have a result for
arbitrary $n$ we can deduce an analytic expression for the leading large $\Nf$
part of the three loop term of 
$\gamma^{\mbox{\footnotesize{$\MSbar$}}}_{\bar{\psi} \sigma^{\mu\nu} D^\rho 
\psi}(a)$ and check that it agrees with our result for $n$ $=$ $2$. In 
particular if we formally write  
\begin{equation} 
\gamma^{(n) \, \mbox{\footnotesize{$\MSbar$}}}_{\bar{\psi} \sigma^{\mu\nu} 
D^\rho \psi}(a) ~=~ C_F \left[ b_1 a ~+~ \left( b_{21} T_F \Nf + b_{20} 
\right) a^2 ~+~ \sum_{r=3}^\infty \sum_{j=0}^{r-1} b_{rj} T_F^j \Nf^j a^r 
\right] 
\end{equation} 
where the coefficients, $\{ b_{ij} \}$, are functions of $n$ and $b_1$, 
$b_{20}$ and $b_{21}$ are known, \cite{38,39,40,37,41}, then we find the 
leading order three loop coefficient is  
\begin{equation}
b_{32} ~=~ \frac{4}{27} \left[ 48 S_3(n) - 80 S_2(n) - 16 S_1(n) 
+ \frac{3( 17 n^2 + 17 n - 8 )}{n(n+1)} \right]  
\end{equation}
where $S_l(n)$ $=$ $\sum_{i=1}^n 1/i^l$. Setting $n$ $=$ $2$ we find exact
agreement with the corresponding term of (\ref{gamtramsb}). Moreover, we can 
expand the critical exponent to four loops and deduce, using the four loop 
$\beta$-function, \cite{46}, 
\begin{eqnarray}
b_{43} &=& \frac{8}{81} \left[ 96 S_4(n) - 160 S_3(n) - 32 S_2(n) 
+ 192 \zeta(3) S_1(n) - 32 S_1(n) \frac{}{} \right. \nonumber \\
&& \left. ~~~~~-~ 144 \zeta(3) + \frac{( 131 n^4 + 262 n^3 + 211 n^2 - 16 n 
- 48 )}{n^2(n+1)^2} \right] ~.   
\end{eqnarray}

\sect{Discussion.} 

To conclude we have derived expressions for the three loop anomalous dimension
of the second moment of the transversity operator in both the $\MSbar$ and
RI$^\prime$ schemes. The results for the latter will be important in the 
extraction of lattice estimates of the associated matrix elements. Moreover, we
have provided the same information for the usual twist-$2$ flavour non-singlet 
operator. Although we have had to consider various projections of the Green's 
function with the operator inserted in a quark two-point function it ought to 
be possible to extend the present calculations to derive the three loop 
anomalous dimensions for higher moments of the transversity operator in the 
$\MSbar$ scheme. In this case since one is only interested in the divergent 
part and not the finite part of the Green's function, it would not require the 
same use of the projections considered here. This is important since the
algebraic manipulations are slowed for a large number of projections and the
inclusion of more covariant derivatives in the operator itself. However, since
those anomalous dimensions are independent of the gauge parameter, the
algebraic computations would be speeded by considering the Feynman gauge. 

\vspace{1cm}
\noindent 
{\bf Acknowledgements.} The author thanks Dr P.E.L. Rakow and Dr C. McNeile for
valuable discussions.

\end{document}